\documentstyle[prl,preprint,aps,tighten]{revtex}
\begin{document}

\title{Replica Symmetry Breaking Instability in the 2D XY model
in a random field}

\author{Pierre Le Doussal}
\address{Laboratoire de Physique Th{\'e}orique
de l'Ecole Normale Sup{\'e}rieure, 24 Rue Lhomond, F-75231 Paris Cedex,
France\cite{prop}}
\author{Thierry Giamarchi}
\address{Laboratoire de Physique des Solides, Universit{\'e} Paris--Sud,
                   B{\^a}t. 510, 91405 Orsay, France\cite{junk}}

\maketitle

\begin{abstract}

We study the 2D vortex-free XY model in a random field, a model
for randomly pinned flux lines in a plane.
We construct controlled RG recursion relations which allow
for replica symmetry breaking (RSB). The fixed point previously
found by Cardy and Ostlund in the glass phase $T<T_c$
is {\it unstable} to RSB. The susceptibility $\chi$ associated to infinitesimal
RSB perturbation in the high-temperature phase is found
to diverge as $\chi \propto (T-T_c)^{-\gamma}$
when $T \rightarrow T_c^{+}$. This provides analytical evidence
that RSB occurs in finite dimensional models.
The physical consequences for the glass phase are discussed.

\end{abstract}
\narrowtext

One of the most fruitful methods to study disordered systems
is the replica method
\cite{edwards_replica} which allows
to average over disorder by introducing $n$ coupled copies of the system.
Since it reduces the problem to a translationally
invariant system this method has the considerable advantage, in principle,
to allow the use of standard field theory techniques. However the limit $n \to
0$
has to be taken, a rather subtle procedure. Indeed,
as is well known from the large body of works on spin glasses
\cite{binder_spinglass_review} the free energy functional is in
general minimized, in a glass phase,
by a solution which breaks replica symmetry spontaneously,
in the hierarchical manner discovered by Parisi
\cite{parisi_replica_breaking}.
Up to now, however, spontaneous replica symmetry breaking (RSB) has been
clearly demonstrated
only when a saddle point method could be used naturally. This is
the case either in infinite-range or mean field models where the saddle point
method provides the exact solution of the problem and RSB is ubiquitous
\cite{mezard_book},
or, as found more recently \cite{mezard_variational_replica}, in field theories
using a Hartree or Gaussian variational method (GVM) which becomes
exact in the limit of large number of components,
in some as yet unelucidated sense.
As explicitly constructed in both cases
the physics of RSB corresponds to systems breaking up in many ``states''
\cite{mezard_book}. Extensions of RSB to finite dimensional
systems
for which such mean-field theory is not exact have been attempted
\cite{dominicis_rsb_fieldtheory}. However, it is far from
obvious, as argued in \cite{fisher_huse_droplets_pure},
that it is relevant in real, i.e low dimensional physical systems.

Another extensively tried method for these problems is the
renormalization group (RG). There
replicas have been used so far simply as
a graph counting trick, to eliminate disconnected graphs in perturbation
theory \cite{lubensky_houches_review}.
This procedure, however, implicitly assumes replica symmetry before
taking the limit $n \to 0$. Although the RG treats fluctuations exactly
and is therefore more accurate
than mean field theory, there is a risk that
it will miss the physics associated to RSB.
Thus one would like to link the two methods.

A good model to look for a renormalization
group that allows RSB is the two dimensional XY model
in a random field
and without vortices (i.e with spins
$S(x)=e^{i \psi(x)}$ where $\psi(x)$ is real-valued in $[-\infty, \infty]$).
This model is particularly interesting
because it is one of the simplest and non trivial models of a ``glass'' in
finite
dimension, to which several analytical methods can be applied
\cite{villain_cosine_realrg,cardy_desordre_rg,%
goldschmidt_dynamics_flux,balents_fermions_0,giamarchi_vortex_short,%
korshunov_variational_short}.
It is relevant to describe several physical disordered systems,
such as randomly pinned flux arrays confined to a plane
\cite{fisher_vortexglass_short,nattermann_flux_creep,giamarchi_vortex_short},
the surface of
crystals with quenched bulk or substrate disorder
\cite{toner_log_2}, planar Josephson junctions
\cite{vinokur_josephson_short},
domain walls in incommensurate solids.
In a pioneering work \cite{cardy_desordre_rg},
Cardy and Ostlund (CO) studied this model using the
renormalization group. They used
the replica method and introduced $n$ coupled XY models
which they mapped onto a Coulomb gas with $n(n-1)/2$ types of vector charges.
They constructed the corresponding RG equations and
performed the limit $n \rightarrow 0$ in these equations, thus
implicitly assuming replica symmetry. The resulting
RG recursion relations, valid near $T_c$, possess a non-trivial
perturbative fixed point for $T<T_c$ at weak disorder strength $g=g^* \propto
T_c-T$.
CO concluded that a ``glass'' phase exists,
controlled by this new fixed point. In this phase,
one coupling constant is found to runaway
to infinity, a somewhat peculiar feature.
These results were extended
in \cite{toner_log_2,hwa_fisher_flux},
and the disorder averaged correlation function
$C(x)= \overline{\langle \psi(x)-\psi(0)\rangle ^2 }$ was
found to grow as $C(x) \sim B {(\log|x|)}^2$, faster than
$C(x) \sim T \log|x|$ which holds
in the high temperature phase and for the pure system.

Although there is presently agreement that
a glass phase exists in this model
for $T < T_c$ its physical properties remain
controversial, despite the large number of analytical and numerical studies.
Two recent numerical simulations
\cite{batrouni_numerical_cardy,cule_numerical_cardy}
have shown somewhat
puzzling results which shed some doubt on the
Cardy Ostlund RG. Both simulations were found incompatible with the
$C(x) \sim B {(\log|x|)}^2$
prediction. In the dynamics \cite{batrouni_numerical_cardy} it was found
that the velocity develops a non linear dependence on the driving
force for $T<T_c$, providing evidence of a glassy phase, but with an
exponent
inconsistent
with the prediction of a
conventional dynamic RG study
\cite{goldschmidt_dynamics_flux}.

These simulations, however, seem compatible with our previous
analytical results using the Gaussian variational method
\cite{giamarchi_vortex_short,korshunov_variational_short}.
We found
that a one-step replica symmetry broken solution described
the glass phase $T<T_c$ with $C(x) \sim T_c \log|x|$ at variance with the
result
of CO, while $C(x) \sim T \log|x|$ for $T>T_c$. Such a discontinuity
in the slope was observed
in \onlinecite{cule_numerical_cardy}. The simulation
\cite{batrouni_numerical_cardy} was
performed at such weak disorder that the system
size was shorter than the length
predicted by the GVM \cite{giamarchi_vortex_short}
beyond which glassy behavior can be
observed in static quantities.
One can argue that a variational method,
such as the GVM,  is approximate even in pure systems
and misses some of the effects of non linearities
better captured by the renormalization group.
On the other hand,  this solution contains the feature of RSB
and its compatibility with numerical calculations suggests
that it has some relevance for
the physics. Thus, it would be quite interesting to have a
direct evidence, in the present model, that spontaneous RSB do occur.

In this paper we present such an evidence. We construct renormalization group
recursion relations,
for the random field XY model, for couplings between replicas of arbitrary
symmetry.
It contains the Cardy Ostlund recursion relations as a particular case. It
allows for a more general way of taking the limit $n \rightarrow 0$, using
Parisi's type matrices, while remaining in the perturbative regime
of the renormalization group. The new operators
introduced in the limit $n \rightarrow 0$ are marginally relevant
at $T=T_c$ and thus can be incorporated in the framework of RG.
We show that the low temperature phase does exhibit spontaneous RSB.
In the high-temperature phase where the RG flows to weak coupling and is
therefore certainly correct, we compute the linear response to a small
RSB perturbation. The associated susceptibility diverges at
the transition as $\chi \propto (T-T_c)^{-\gamma}$
when $T \rightarrow T_c^{+}$. For $T<T_c$ the replica symmetric flow
and the CO fixed point, are unstable to a small RSB. This is to our knowledge
the first physical model where this effect can be demonstrated in a controlled
way.

The Hamiltonian of the 2D vortex-free XY model
in a field of random amplitude
and direction, reads:
\begin{eqnarray} \label{2DXY}
H &=& \int d^2x\;\frac{c}{2} (\nabla \psi(x))^2
- \eta(x) \cdot \nabla \psi(x)  \\
& & - \zeta_1(x) \cos(\psi(x))
- \zeta_2(x) \sin(\psi(x)) \nonumber
\end{eqnarray}
with $\overline{ \eta_i(x) \eta_j(x') }=\Delta_0 \delta_{ij} \delta(x-x')$
and $\overline{ \zeta_i(x) \zeta_j(x') }=\Delta \delta_{ij} \delta(x-x')$
are two Gaussian white noises. This model also
describes flux lines with displacements $u$ and average spacing $a$ in $d=1+1$
dimensions,
with $\psi = 2 \pi u/a$.
$\Delta$ is proportional to the amplitude of disorder with Fourier
component close to $2 \pi/a$ and $\Delta_0$ is the long wavelength
disorder. Note that $\Delta_0$ is generated if not present originally in
the model.

After replication of (\ref{2DXY}) and averaging one obtains:
\widetext
\begin{equation} \label{generalham}
\frac{H_{\text{eff}}}{T} =  \int d^2x  \sum_{ab} \left[
 \frac{K_{ab}^{-1}}{2\pi} \nabla \phi^a \nabla \phi^b
- \frac{g_{ab}}{(2\pi)^2} \cos(2(\phi^{a}(x)-\phi^{b}(x)) \right]
\end{equation}
\narrowtext
where we set $g_{aa}=0$. We have used for convenience:
\begin{equation}  \label{identification}
K_{ab}^{-1} =  \frac{4\pi c}{T} \delta_{ab} -
                  \frac{4\pi\Delta_0}{T^2} , \qquad
\frac{g_{ab}}{(2\pi)^2}  =  \frac {\Delta}{2 T^2}
\end{equation}
and
$\phi^{a}  =  \psi^{a}/2 $.
This defines the ``bare'' or starting values parameters, which
of course are replica symmetric.
We now consider a more general situation
where the``renormalized'' parameters $K_{ab}$ and $g_{ab}$
have arbitrary symmetry with respect to the group of
replica permutation. We use the parameterization:
\begin{equation}\label{kdef}
K^{-1}_{ab} = \delta_{ab} - k_{ab}
\end{equation}
One defines the connected part $k_c = \sum_b k_{ab}$.
As will be obvious later there is a transition at temperature
$T_c = 4\pi c$ in the model (\ref{2DXY}). Using (\ref{identification})
and (\ref{kdef}) this corresponds to $k_c=0$, and one has more generally
$k_c = \frac{T-T_c}{T} = \tau$, where $\tau$ is the reduced temperature.

In previous applications of RG to disordered systems, recursion relations
are established for arbitrary number $n$ of replica. Then replica symmetry is
assumed and the RG equations become simple functions of $n$.
The continuation
$n \rightarrow 0$ is then easily taken. In that respect
the use of replica is a trick of graph counting. In fact
one can generally establish identical RG equations
directly by considering disorder propagators, a method
which we call ``replica symmetric perturbation theory''. However, in a
glassy system, where many metastable states exist, one can question the
validity
of such a renormalization group procedure.
A crucial assumption of the RG is that one can
simply integrate over the short scale degrees of freedom
independently of the larger scales, to produce an equivalent renormalized
Hamiltonian. It is not obvious that
such a separation between scales exists in glassy systems.
Short scale degrees of freedom may well be determined by
the local minimum they belong to, thus depend in effect on larger scales.
The presence of long range effects is
supported by the massless modes found in the
expansions around Parisi's solution
\cite{dominicis_rsb_fieldtheory}. However,
the replicated Hamiltonian is translationally invariant and
all these problems are
buried in the proper taking of the limit $n \to 0$.
In the case where the standard replica symmetric limit $n \to 0$ does not work,
a Parisi type RSB might allow to construct a correct RG for glassy
systems. We will therefore construct RG equations keeping the
full matrix structure of the couplings. Since we are looking for a scale
invariant
 theory
{\it near the transition}, we consider perturbation
theory in $\tau$, in
the small matrix parameters $2 k_{ab}-k_{aa}-k_{bb}$ and in $g_{ab}$.
This leaves unconstrained
the degree of freedom of a
constant shift of all the elements of $k_{ab}$.
$K^{-1}_{ab}$ is easily inverted as
$K_{ab} = (1-k_c)^{-2}[(1-2k_c) \delta_{ab} + k_{ab}]
       \simeq \delta_{ab} + k_{ab}$,
which is valid even if the $k_{ab}$ are large provided
the above-mentioned parameters are small.

The constants $g_{ab}$, $k_{ab}$ and $k_c= \tau$ correspond
to quadratic interactions in an
equivalent fermion problem \cite{solyom_revue_1d}.
Using standard renormalization group
methods, either on the fermion form or
on the classical Hamiltonian (\ref{generalham})
\cite{solyom_revue_1d}, one can derive to second order
the general RG for the replicated system
\begin{eqnarray} \label{lessol}
\frac{d g_{ab}}{dl}  & = &  (2k_{ab}-k_{aa}-k_{bb})g_{ab}
                     +  \frac{1}{2} \sum_{c\ne a,b} g_{ac} g_{bc}
                        \nonumber \\
\frac{d k_{ab}}{dl}  & = &   \frac{1}{4} g_{ab}^2 , \qquad
\frac{d \tau}{dl}  =  0
\end{eqnarray}
where $l$ is the standard logarithmic scale.
In these equations $a \ne b$ is implied.
Note that the reduced temperature $\tau$ is unrenormalized.
The subspace of replica-symmetric parameters $g_{ab}=g$, $k_{a\ne b}=k$
is obviously preserved by the RG and the equations (\ref{lessol}) for
$g$, $k$ and $\tau$ reduce to the one obtained by Cardy and Ostlund
in the limit $n=0$:
\begin{equation}
\frac{d g}{dl}   =   - 2 \tau g - g^2 , \qquad
\frac{d k}{dl}   =   \frac{1}{4}  g^2
\end{equation}
In the high temperature phase, $\tau >0$, the disorder $g$ is irrelevant
and the fixed point is a pure Gaussian system with
$C(x)= \overline{\langle \phi(x)-\phi(0)\rangle ^2 }
\sim (\frac{T}{T_c} + O(g) ) \log|x|$. The renormalization of
the off-diagonal part $k$ contributes to $O(g)$ since
$k$ goes to a constant (which stays
finite up to $\tau=0$).
In the low temperature phase, there is a fixed point at $g^*= 2 |\tau|$.
Note however that $k$ flows to infinity. This is a peculiar situation which
within the replica symmetric scheme does not lead to inconsistencies,
since $k$ does not feedback to any order
in perturbation theory (only averages
$\overline{(\sum_a C_a \phi_a)^2}$ with $\sum_a C_a=0$ appear).
The flow of $k$ was used \cite{toner_log_2} to predict $C(x) \sim B
{(\log|x|)}^2$.

Let us now look at the RG flow for coupling constants parameterized
by Parisi matrices. $g_{ab}$ and $k_{ab}$, $a \ne b$
are replaced by functions $g(u)$, $k(u)$, with $0<u<1$.
The symmetric case corresponds to constant functions.
The independent variables are now
$\tau$, $g(u)$ and $k(u)$.
The equations now read:
\widetext
\begin{equation} \label{lessol2}
\frac{d g(u)}{dl}  =
(- 2 \tau  + 2(k(u) - \langle k \rangle) ) g(u)
- \langle g \rangle g(u)
       - \frac{1}{2} \int_0^u dv (g(v)-g(u))^2 , \qquad
\frac{d k(u)}{dl}  =  \frac{1}{4} g(u)^2
\end{equation}
\narrowtext
where $\langle k\rangle =\int_0^1
dv k(v)$. These functions depend
on $l$ but this will be written explicitly only when needed.
Some features are immediately apparent on (\ref{lessol2}).
There is now a spectrum of dimensions,
given by $- \tau + k(u) - \langle k \rangle$,
for the operators corresponding to the non replica symmetric
couplings $g_{ab}$. The dimension of these operators should be small
to neglect the effect of higher order replica terms. Also $k(u)$
now feeds back in the equation for $g(u)$,
thus one anticipates an instability.

Let us first check the stability of the replica symmetric flow
in the low temperature phase. We separate the replica symmetric part and
write $g(u)=g_l + \epsilon(u)$ and
$k(u)= k_l + A(u)$ with $\langle \epsilon \rangle=0$ and
 $\langle A \rangle=0$. Starting from the CO fixed point
$g=g^*$, $k_l$ being the running coupling constant when $g=g^*$,
one obtains from (\ref{lessol2}) the deviations from
the replica symmetric solution to linear order:
\begin{equation} \label{lessol3}
\frac{d\epsilon(u)}{dl}  =  2 g^* A(u) , \qquad
\frac{dA(u)}{dl}  =  \frac{g^*}{2} \epsilon(u)
\end{equation}
Thus the RG flow is clearly {\it unstable} to replica
symmetry breaking when $\tau < 0$. The eigenvalue
of instability is $\lambda=2 |\tau|$.
When $\tau<0$ there is no small coupling fixed point and
the flow goes to strong coupling \cite{footnote2}. This is clear from
(\ref{lessol2}) since $A(u)$ can only reach
a fixed point
if $g^2(u) - \langle g^2 \rangle$ goes to zero.
This is a strong indication that the low temperature phase
corresponds to a replica symmetry broken solution.
To conclude unambiguously on that issue, we now consider
the high temperature phase $\tau > 0$ where the RG is exact
since there is a weak coupling fixed point
$g(u) = 0$ (and $A(u)$ small).
By computing the linear susceptibility to small RSB perturbation
we show that spontaneous RSB occurs at $T_c$.

Let us define the susceptibility to RSB as follows.
We start, for $\tau >0$ with a given disorder $g_0$ and add a small
RSB perturbation $g(u)=g_0 + \epsilon_0(u)$ uniform in space.
This will result in replica non symmetric part in the correlation
functions $\langle \phi_a(q) \phi_b(-q) \rangle = (\delta_{ab} + k^*_{ab})/q^2$
where $k^*_{ab}$ are the renormalized coupling at the fixed point $g(u)=0$.
A susceptibility is defined as:
\begin{equation}\label{chidef}
A^*(u) \equiv k^*(u) - \langle k^* \rangle = \chi \epsilon_0(u)
\end{equation}
in the limit $\epsilon_0 \to 0$. One should consider a $\chi$
function of $u$ but it turns out that it has the above simple form.
Alternatively, one can add a quadratic RSB perturbation
$\delta k_{ab} \nabla \phi_a \nabla \phi_b$ and define a corresponding
susceptibility $\chi'$ as in (\ref{chidef}).
To compute $\chi$ let us expand the RG equations
(\ref{lessol2}) to linear order in $\epsilon_0$. As can be seen from
(\ref{lessol2}), $A$ can be considered as linear
in $\epsilon_0$.
\begin{eqnarray} \label{lesbase}
\frac{d g}{dl}  & = & - 2 \tau g - g^2
       , \qquad \frac{d A(u)}{dl} = \frac{1}{2} g \epsilon(u) \nonumber \\
\frac{d \epsilon(u)}{dl}  & = & - ( 2 \tau + g) \epsilon(u) + 2 g A(u)
\end{eqnarray}
At linear order, each $u$ can be treated independently.
We perform the following rescalings $x=2 \tau l$, $\epsilon(u)=
\epsilon_0(u) f$, $A(u) = \epsilon_0(u) a$, $g=2 \tau \tilde{g}$.
The first equation of (\ref{lesbase}) integrates to give
$ \tilde{g}(x)= \frac{ e^{-x} }{ \alpha + 1 - e^{-x} }$
where we have defined $\alpha= 2 \tau/g_0$. Defining
$Z(x)= \frac{2}{\alpha} \int_0^x dy A(y) + 1$,
(\ref{lesbase}) gives
$f(x)= \alpha e^{-x} Z(x)/(\alpha + 1 - e^{-x} )$
where $Z(x)$ satisfies:
\begin{equation}\label{decadix}
\frac{d^2 Z}{dx^2} = Z(x) e^{-2 x}/(\alpha + 1 - e^{-x})^2
\end{equation}
with initial conditions, $Z(0)=1$ and $Z'(0)=0$. The problem depends
thus only on one variable $\alpha=2 \tau/g_0$. The susceptibility
is obtained from the asymptotic value $A_{\infty}$ and
is $\chi = \lim_{x \to \infty} \alpha Z'(x)/2$.

A full solution of (\ref{decadix}) can be obtained numerically, but
$\chi$ can be estimated from considering the two asymptotic regimes
$\alpha \gg 1$ and $\alpha \ll 1$. In the first regime
$\alpha \gg 1$, (\ref{decadix})
becomes $Z''(x)= e^{-2 x} Z(x)/\alpha^2$, whose solution is in terms
of Bessel functions,
$Z(x)= c_1 I_0(e^{-x}/\alpha) + c_2 K_0(e^{-x}/\alpha)$ . $c_{1,2}$
are two constants determined from the initial conditions. This gives
in that regime:
\begin{equation}\label{debug}
\chi = \frac{1}{2} I_1(\frac{g_0}{2 \tau}) \sim \frac{g_0}{8 \tau}
\end{equation}
To investigate the behavior close to the transition
one has to look at $\alpha \ll 1$ and
to match two regimes in $x$. For $x \ll 1$,
(\ref{decadix}) can be written $Z''(x) = Z(x)/(\alpha + x )^2$
whose solution are power laws $Z \propto (\alpha +x )^{-\nu}$
where $\nu$ satisfies the golden mean equation $\nu^2 + \nu =1$.
We define $\gamma=(\sqrt{5}-1)/2 \simeq 0.618$. The solution
is $Z(x)= ((1+x/\alpha)^{-\gamma} +\gamma^2
(1+x/\alpha)^{1/\gamma})/(1+\gamma^2)$.
When $x$ becomes of order $1$,
one can use the equation of the first regime
with $\alpha \to \alpha +1 $. A simple matching at $x \approx 1$ gives:
\begin{equation}\label{diverge}
\chi \propto  C (g_0/2 \tau)^{\gamma}
\end{equation}
We performed a numerical integration of (\ref{decadix}) which confirms
the analytic estimates (\ref{debug},\ref{diverge}) and gives
$C \approx 0.165$.
$\chi'$ can be computed similarly
from (\ref{decadix}), with modified initial conditions $Z(0)=0$ and
$Z'(0)=2/\alpha$
and has the same divergence as $\chi$.
There is also a non linear response in $\epsilon_0$ in the high temperature
phase. A runaway flow occurs
for smaller and smaller values of $\epsilon_0$
when $\tau \to 0$, roughly when $\chi \epsilon_0 > \tau$.

The divergence of $\chi$ in (\ref{diverge}) when $\tau \to 0$
fixed $g_0$ shows that RSB occurs spontaneously \cite{footnote}
for $\tau \leq 0$. Quantities like $\langle \phi_a(q) \phi_b(-q) \rangle$
acquire a replica non symmetric part. Defining
$q_{ab}=\langle \nabla \phi_a(x) \nabla \phi_b(x) \rangle$, a possible order
parameter for RSB is
$Q_{ab} = 2 q_{ab} + \sum_{c \ne a} q_{ac} + \sum_{c \ne b} q_{bc}$. The
perturbation
$\delta k_{ab}$ is the conjugated field to $Q_{ab}$. Thus $\chi'$ is
the inverse effective mass of $Q_{ab}$, and its divergence signals
an instability \`a la de Almeida Thouless \cite{dealmeida_thouless}
when $T \leq T_c$.
By analogy with the sine Gordon model, a reasonable interpretation
of the runaway flow of $g_{ab}$ is that a ``mass'' develops for some
modes below $T_c$.
This scenario is in good agreement with the findings of the variational
method\cite{giamarchi_vortex_short,korshunov_variational_short},
where correlation functions become non replica symmetric
and a fraction $1-u_c$ of the modes become massive below $T_c$,
with $u_c \approx T/T_c$.
These modes correspond, in the Coulomb gas, to
types of charges which unbind. Note that
a fraction $u_c$ remains massless (corresponding to bound types
of charges).
This method, being non perturbative contrary to RG,
allows for the generation of a mass.

In this paper, we have shown that replica symmetry breaking can be incorporated
in the RG. We found new relevant operators
which break replica symmetry when $n \to 0$ for $T \leq T_c$. The associated
susceptibility diverges in the high temperature phase $T > T_c$.
The replica symmetric fixed point found by Cardy
Ostlund is unstable. Observable consequences of RSB could be looked for in
the dynamics of this system. RSB is usually
accompanied by
effects of aging, persistent correlations,
and breaking of fluctuation dissipation theorem
\cite{dynamics}.
In our opinion it is also probable that the static correlation functions
should be different from the one predicted by the symmetric RG, as hinted at
by the variational method. Further numerical and experimental results
would be of great interest.

We are grateful to E. Brezin, M. Gabay, T. Hwa and Y. Shapir
for useful discussions.


\begin{thebibliography}{10}
\bibitem{footnote2}
A RSB coupling fixed point perturbative in $g_{ab}$ could
be accessible if higher orders in $K_{ab}$ are included.
\end{thebibliography}

\end{document}